\documentclass[12pt]{article}
\usepackage{amsfonts}
\makeatletter
\@addtoreset{equation}{section}

\makeatother

\begin{document}
\begin{titlepage}
 
$\mbox{ }$
\begin{flushright}
\begin{tabular}{l}
KUNS-1796\\
NEIP-02-006\\
hep-th/0209057\\
Sep 2002
\end{tabular}
\end{flushright}

~~\\
~~\\
~~\\

\vspace*{0cm}
    \begin{Large}
       \vspace{2cm}
       \begin{center}
        Curved-space classical solutions of a massive supermatrix model
       \end{center}
    \end{Large}

  \vspace{1cm}

\begin{center}
          Takehiro A{\sc zuma}$^{\dag}$\footnote{
e-mail address : azuma@gauge.scphys.kyoto-u.ac.jp} and
          Maxime  B{\sc agnoud}$^{\ddag}$\footnote
           {
e-mail address : Maxime.Bagnoud@unine.ch} \\

\vspace*{1cm}
          $^{\dag}${\it Department of Physics, Kyoto University,
Kyoto 606-8502, Japan}\\
\vspace*{1cm}
       $^{\ddag}$ {\it Institut de Physique, Universit\'e de Neuch\^atel,
   CH-2000 Neuch\^atel, Switzerland}

\end{center}

\vfill

\begin{abstract}
\noindent
We investigate here a supermatrix model with a mass term and a cubic
interaction. It is based on the Lie superalgebra 
$\mathfrak{osp}(1|32,\mathbb{R})$, which could play a r\^ole in the 
construction of the eleven-dimensional M-theory. 
This model contains a massive version of the IIB matrix model, where
some fields have a tachyonic mass term. Therefore, the trivial 
vacuum of this theory is unstable. However, this model possesses several
classical solutions where these fields build noncommutative curved spaces 
and these solutions are shown to be energetically more favorable than 
the trivial vacuum. In particular, we describe in details two cases, 
the $SO(3) \times SO(3) \times SO(3)$ (three fuzzy 2-spheres) and the 
$SO(9)$ (fuzzy 8-sphere) classical backgrounds.
\end{abstract}
\vfill
\end{titlepage}
\vfil\eject

\section{Introduction}
  Despite the fact that perturbative superstring theory provides us with a 
  consistent unified theory of fundamental interactions, we still lack a
  completely satisfactory nonperturbative formulation of superstring 
  theory. 
  As a consequence, although this theory has a plethora of possible vacua
  (the dynamics in some of these vacua has already been studied in details),
  there is no way to select the true vacuum of the theory and compare
  the physical implications of superstring theory with known phenomenological 
  data. It is thus instrumental to find a constructive definition of 
  superstring theory in order to predict the real world or/and
  falsify the theory.

  One of the successful proposals for a constructive definition of
  superstring theory\cite{9610043,9612115,9703030,9708123}
  is a formulation through a large $N$ reduced model. A candidate model
  of this kind is the so-called IIB matrix model\cite{9612115,9705128}, 
  which is defined by the following action:\footnote{We draw the readers'
  attention to our choice of sign for the bosonic term of the IIB matrix 
  model. In this paper, we regard the action as {\it minus} the potential, 
  which is a choice of sign opposite to the usual definition.
  We define the action (\ref{AZ1IKKT}) in the 10-dimensional
  Minkowskian space, and the path integral is in our case defined as
  \begin{eqnarray}
   Z = \int dA d \psi e^{+S_{E}}, \nonumber 
  \end{eqnarray}
 where $S_{E}$ is defined in the 10-dimensional Euclidean space;
 i.e. $S_{E}$ is defined by Wick-rotating $A_{0}$ as $A_{0} \to i A_{0}$
 and replacing the gamma matrices for the $SO(9,1)$ Clifford algebra with 
 those for $SO(10)$ in the action $S$.} 
  \begin{eqnarray}
  S = \frac{1}{g^{2}} Tr \left( \frac{1}{4} [A_{\mu},
  A_{\nu}][A^{\mu}, A^{\nu}] + \frac{1}{2} {\bar \psi} \Gamma^{\mu}
  [A_{\mu}, \psi] \right), \label{AZ1IKKT}
  \end{eqnarray}
 where the indices $\mu, \nu, \cdots$ run over the 10-dimensional
 Minkowskian spacetime. 
 It is a large $N$ reduced model\cite{EguchiKawai,TEK,TEK2} of 10-dimensional
 ${\cal N} =1$ super Yang-Mills theory with $U(N)$ gauge symmetry. Here, 
 $\psi$ is a 10-dimensional Majorana-Weyl spinor field, and $A_{\mu}$ and 
 $\psi$ are $N \times N$ Hermitian matrices. The IIB matrix model has 
 ${\cal N}=2$ supersymmetry, which exhibits a particular structure that 
 allows us to interpret the eigenvalues of the large $N$ matrices describing 
 the bosonic fields as space-time coordinates\cite{9802085,9903217} 
 (IIB matrix model is extensively reviewed in \cite{9908038}). 

 Another intriguing attempt for a constructive definition of superstring 
 theory is a background-independent matrix model based on the Lie 
 superalgebra $\mathfrak{osp}(1|32,\mathbb{R})
 $\cite{0002009,0006137,0102168,0103003,0201183}.
 It is a natural generalization of the IIB matrix model, in which 
 both bosons and fermions are unified into a single supermultiplet.  
 $\mathfrak{osp}(1|32,\mathbb{R})$ has been known as the unique maximal 
 simple Lie superalgebra with 32 fermionic generators\cite{0003261}. 
 In a 10-dimensional representation, the smallest irreducible
 spinors are the 16-component chiral spinors, so that the 32 fermionic
 generators can be decomposed in two chiral spinors of equal or opposite
 chiralities. The former and the latter respectively correspond to the type 
 IIA and IIB superstring theories.
 In this sense, we speculate that the IIB matrix model would be extracted from
 this supermatrix model by integrating out some degrees of freedom.
 In the papers \cite{0102168,0103003}, it has been attempted to clarify 
 the relation between a purely cubic $\mathfrak{osp}(1|32,\mathbb{R})$ 
 supermatrix model and the IIB matrix model by paying particular attention
 to the structure of the supersymmetry algebra. In the paper \cite{0201183},  
 such a cubic supermatrix model supplemented by a mass term has been 
 investigated to elucidate its relation with the BFSS matrix 
 model\cite{9610043}. 

 Since large $N$ reduced models are expected to be eligible
 frameworks to describe gravitational interactions, it is essential
 to have the possibility of describing curved spacetimes manifestly in 
 their framework. As the IIB matrix model only possesses flat noncommutative 
 spacetime as a classical solution (the same holds true of the non-gauged
 $\mathfrak{osp}(1|32,\mathbb{R})$ supermatrix model\cite{0102168}), 
 it is impossible to study perturbations around curved backgrounds.

 Some generalization is thus necessary in order to overcome this difficulty. 
 A possible approach is to identify large $N$ matrices with differential 
 operators\cite{0102168,0201129,0202138,0204078}. 
 Large $N$ matrices have both aspects of differential operators and 
 spacetime coordinates. The former appears clearly in the twisted Eguchi-Kawai 
 model\cite{TEK,TEK2} while the latter is the essential feature of the
 IIB matrix model. These two aspects can be related by expanding the 
 IIB matrix model around its flat noncommutative background\cite{9908141}. 
 This bilateral character is interpreted as the T-duality of string theory. 
 The advantage of identifying matrices with differential operators lies 
 in the fact that differential operators act on fields on a curved spacetime
 in a natural way. 

 Another approach is to consider a matrix model which has some curved space 
 as a classical solution, so that it becomes possible to perform perturbations 
 around this curved background. To achieve this, some modification of the IIB 
 matrix model\cite{0101102,0103192,0204256,0207115} is needed.
 In \cite{0101102}, a Chern-Simons term has been added
 to the IIB matrix action to construct a noncommutative gauge theory 
 on the $SO(3)$ fuzzy sphere\cite{Madore}.  
 Another possible alteration is the addition of a tachyonic mass term 
 to the bosonic part of the IIB matrix model\cite{0103192}:
  \begin{eqnarray}\label{MIIBMM}
   S = \frac{1}{g^{2}} Tr \left( \frac{1}{4} [A_{\alpha}, A_{\beta}] 
   [A_{\alpha}, A_{\beta}] + \lambda^{2} A_{\alpha} A_{\alpha} \right),
  \end{eqnarray}
 where the indices $\alpha, \beta$ run over $\alpha, \beta = 1, 2, 3, 
 4$. The indices are contracted with respect to the four-dimensional 
 Euclidean space metric, and the model has $SO(4)$ global symmetry.
 Its equations of motion
  \begin{eqnarray}
   [A_{\beta}, [A_{\alpha}, A_{\beta}]] + 2 \lambda^{2} A_{\alpha} = 0 
  \end{eqnarray} 
 have classical solutions given by a set of fields satisfying some 
 Lie algebra.  Thus, such a massive IIB matrix model can be 
 expanded around various curved spaces. In \cite{0103192}, expansions 
 around the two-dimensional fuzzy sphere and the two-dimensional fuzzy 
 torus have been studied.

 In this paper, we take this latter approach in order to describe a curved 
 background spacetime by considering an $\mathfrak{osp}(1|32,\mathbb{R})$ 
 supermatrix model with a mass term. We analyze how the massive supermatrix 
 model incorporates the non-commutative curved-space classical solutions.  

  This paper is organized as follows:
  In Section 2, we give a brief review of the Lie superalgebra
  $\mathfrak{osp}(1|32,\mathbb{R})$ and the associated
  supermatrix model. 
  In Section 3, we suggest an ansatz that allows us to solve the equations 
  of motion of the massive supermatrix model and leads to solutions
  of the fuzzy-sphere type.  We describe in detail two of these solutions, 
  one exhibiting $SO(3) \times SO(3) \times SO(3)$ symmetry and the other 
  exhibiting $SO(9)$ symmetry and compare their stability properties. 
  This leads us to a more general discussion of a possible brane nucleation 
  process in such totally reduced matrix models. Then, we make a few remarks 
  on the structure of the supersymmetry transformations in our model. 
  Finally, we summarize the results presented in this work 
  in section 4 and indicate there a few directions for future research 
  on this topic.
 
 \section{$\mathfrak{osp}(1|32,\mathbb{R})$ supermatrix model with a mass term}
  
  L. Smolin proposed a cubic matrix model\cite{0002009,0006137} based
  on the Lie superalgebra $\mathfrak{osp}(1|32,\mathbb{R})$. The action is 
  constructed from a matrix $M$ belonging to 
  $\mathfrak{osp}(1|32,\mathbb{R})$, whose entries are promoted to large 
  $N$ Hermitian matrices. 
  In this paper, we follow the notation of \cite{0103003}, in which details
  about $\mathfrak{osp}(1|32,\mathbb{R})$ are given. Its relation with the 
  11-dimensional super-Poincar\'e algebra is described in 
  \cite{0003261,0201183}. 
 
  The (even part) of the Lie superalgebra $\mathfrak{osp}(1|32,\mathbb{R})$ is
  constituted from matrices of the form:
  \begin{eqnarray}
    M = \left( \begin{array}{cc} m & \psi \\ i {\bar \psi} & 0
    \end{array} \right), 
   \end{eqnarray}
 where $\psi$ is a 32-component Majorana spinor and $m$ belongs 
 to the Lie algebra $\mathfrak{sp}(32,\mathbb{R})$. We take the  metric of 
 the 10-dimensional Minkowskian spacetime as 
  \begin{eqnarray}
 \eta^{\mu \nu} = \textrm{diag}(-1, +1, \cdots, +1).
  \end{eqnarray}

 \subsection{Action} 
  Here, we consider an $\mathfrak{osp}(1|32,\mathbb{R})$ supermatrix model 
  with a mass term included, expecting similarities with the massive IIB
  matrix model studied in \cite{0103192}.
  We consider the following action, with a mass term added to the
  pure cubic $\mathfrak{osp}(1|32,\mathbb{R})$ supermatrix model:
     \begin{eqnarray}
       S &=& Tr_{\mathfrak{u}(N)} \left[ str_{\mathfrak{osp}(1|32)} \left( 
       - 3 \mu M^{2} + i M [M, M]
       \right) \right]  \nonumber \\
       &=& Tr_{\mathfrak{u}(N)} \left[ 3 \mu (- {m_{p}}^{q}{m_{q}}^{p}
       + 2i {\bar \psi} \psi) + i \left( {m_{p}}^{q}
        [ {m_{q}}^{r}, {m_{r}}^{p} ] - 3 i {\bar \psi}^{p}
        [ {m_{p}}^{q}, \psi^{q}] \right) \right].  \label{AZaction}
   \end{eqnarray}
 where $p, q, r, \cdots = 1,\ldots, 32$.
 In this model, each element of the $\mathfrak{osp}(1|32,\mathbb{R})$ 
 supermatrices is promoted to an $N \times N$ Hermitian matrix.
 This action is invariant under $U(N)$ gauge transformations and 
 $OSp(1|32,\mathbb{R})$ orthosymplectic transformations, and 
 these two symmetries are decoupled, since they do not act on the same 
 indices. Since we want to consider this model in a 10-dimensional spacetime 
 context, we decompose the bosonic part $m$ as follows: 
  \begin{eqnarray}
   m = W \Gamma^{\sharp} + A_{\mu} \Gamma^{\mu} + B_{\mu} \Gamma^{\mu \sharp}
     + \frac{1}{2!} C_{\mu_{1} \mu_{2}} \Gamma^{\mu_{1} \mu_{2}}
     + \frac{1}{4!} H_{\mu_{1} \cdots \mu_{4}} \Gamma^{\mu_{1} \cdots
     \mu_{4} \sharp}
     + \frac{1}{5!} Z_{\mu_{1} \cdots \mu_{5}} \Gamma^{\mu_{1} \cdots \mu_{5}},
  \end{eqnarray}
 where $\mu_i=0,\ldots ,9$ and $\Gamma^{\sharp}$ is the chirality operator.
  Then, the relevant part of the action (\ref{AZaction}) is expressed as
 (writing simply $Tr$ instead of $Tr_{\mathfrak{u}(N)}$ from now on):
 \begin{eqnarray}
 S &=& 96 \mu Tr \left( - W^{2} - A_{\mu} A^{\mu} + B_{\mu} B^{\mu}
   + \frac{1}{2} C_{\mu_{1} \mu_{2}} C^{\mu_{1} \mu_{2}}
     - \frac{1}{4!} H_{\mu_{1} \cdots \mu_{4}} H^{\mu_{1} \cdots
     \mu_{4}} - \right. \nonumber \\ 
  &-& \left. \frac{1}{5!} Z_{\mu_{1} \cdots \mu_{5}} Z^{\mu_{1} \cdots
     \mu_{5}} +  \frac{i}{16} {\bar \psi} \psi \right)
    + 32 i Tr \Bigg( 3 C_{\mu_{1} \mu_{2}}  [B^{\mu_{1}}, B^{\mu_{2}}] 
     + C_{\mu_{1} \mu_{2}} [ {C^{\mu_{2}}}_{\mu_{3}}, C^{\mu_{3} \mu_{1}}]
    \Bigg) + \nonumber \\ 
  &+& \textrm{ cubic interactions involving } (W,A,H,Z,\psi), \label{AZaction3}
  \end{eqnarray} 
 while the full result can be found in the first appendix and the detailed 
 computation in \cite{0103003}. In the purely cubic supermatrix model 
 (without mass term, which has been studied in 
 \cite{0002009,0006137,0102168,0103003}), the rank-2 field 
 $C_{\mu_{1} \mu_{2}}$ possesses a cubic interaction term but has no 
 quadratic term. This has been a severe obstacle to the appearance of 
 a Yang-Mills-like structure in the supermatrix model, because it has been
 impossible to identify $C_{\mu_{1} \mu_{2}}$ with the commutators of
 the rank-1 fields $[B_{\mu_{1}}, B_{\mu_{2}}]$ (or $[A_{\mu_{1}},
 A_{\mu_{2}}]$). In the 11-dimensional case, this difficulty has been overcome 
 in \cite{0201183} through the addition of a mass term, and we thus expect 
 this model to contain the massive IIB matrix model, the bosonic part of which
 has been studied in \cite{0103192} to investigate perturbation theory around
 noncommutative curved-space backgrounds.

\section{Resolution of the equations of motion}
 We proceed to search for possible curved-space classical configurations
 solving the equations of motion that follow from the action 
 (\ref{AZaction3}). To get a clearer picture of the problem, we now set 
 the fermions and the positive squared-mass bosonic fields to zero: 
  \begin{eqnarray}
   \psi = W = A_{\mu} = H_{\mu_{1} \cdots \mu_{4}} = Z_{\mu_{1} \cdots
   \mu_{5}} = 0. \label{AZ2trivial}
  \end{eqnarray}
 Since their masses are positive (at least in the spatial directions,
 while the time-like direction of quantum fields is generally unphysical), 
 (\ref{AZ2trivial}) is a stable classical solution. 
 Furthermore, we choose to identify the tachyonic 10-dimensional vector field 
 $B_{\mu}$, rather than the well-defined  $A_{\mu}$ with the bosonic fields 
 of the massive IIB matrix model, in order to obtain a possibly stable 
 curved-space classical solution. 
 The classical equations of motion for the remaining tachyonic fields 
  $B_{\mu}$ and $C_{\mu \nu}$ following from (\ref{AZaction3}) are
  \begin{eqnarray}
   B_{\mu} &=& - i \mu^{-1} [B^{\nu}, C_{\mu \nu}], \label{AZeqB+} \\
   C_{\mu \nu} &=& - i \mu^{-1} ( [B_{\mu}, B_{\nu}]
    + [ {C_{\mu}}^{\rho}, C_{\nu \rho}] ). \label{AZeqC+}
  \end{eqnarray}
 Although it is difficult to solve these equations in full generality,
 the equation of motion for $C_{\mu \nu}$ suggests to take 
 $C_{\mu \nu} \propto [B_{\mu}, B_{\nu}]$ for $B_{\mu}$'s satisfying a fairly
 simple commutator algebra. If we look for objects having a clear geometrical 
 interpretation, it is tempting to look for solutions building fuzzy spheres.
 
 \subsection{$SO(3) \times SO(3) \times SO(3)$ classical solution}
 The simplest tentative solution is the product of three fuzzy 2-spheres 
 with the symmetry $SO(3) \times SO(3) \times SO(3)$. Such a system is
 described by $N \times N$ hermitian matrices building a representation
 of the $\mathfrak{so}(3)$ Lie algebra in the following way:
  \begin{equation}
   [B_{i}, B_{j}]  = i \mu r
   \epsilon_{ijk} B_{k}, \hspace{12mm} B_{1}^{2} + B_{2}^{2} + B_{3}^{2} =
   \mu^{2} r^{2} \frac{N^{2} - 1}{4} {\bf 1}_{N \times N} 
   \textrm{ for } (i,j,k=1,2,3)\label{AZso3so3so3} 
  \end{equation}
 with similar relations for $i,j,k=4,5,6$ and $i,j,k=7,8,9$, trivial 
 commutators for indices that do not belong to the same group of 3,
 and $B_{0} = 0$ ($\epsilon_{ijk}$ is defined as usually).

 This set of fields (\ref{AZso3so3so3}) describes a space formed
 by the Cartesian product of three fuzzy spheres located in the directions
 $(x_{1}, x_{2}, x_{3})$, $(x_{4}, x_{5}, x_{6})$ and $(x_{7}, x_{8},
 x_{9})$, whose radii are all $\mu r \sqrt{N^{2}-1}/2$.
 $(N^{2}-1)/4$ is the quadratic Casimir operator of the $\mathfrak{so}(3)$ 
 Lie algebra. 
 Note that any positive-integer value of $N$ is possible here, since
 $N$ indexes the dimensions 
 of irreducible representations. For $SO(3)$, the irreps have dimensions
 $N=2j+1$, for all integer values of the spin $j$. However, we can also use 
 spinorial representations with half-integer spins in this case.   
 We have to consider this classical solution instead of the single $SO(3)$ 
 fuzzy sphere
  \begin{eqnarray}
   [B_{i}, B_{j}] = i \mu r \epsilon_{ijk} B_{k} \textrm{ (for 
   $i,j,k =1,2,3$)}, \hspace{3mm} B_{\mu} = 0 \textrm{ (for $\mu=0, 4,
   5, \cdots, 9$)},
  \end{eqnarray}
 because the solution $B_{4} = \cdots = B_{9} = 0$ is unstable in the 
 directions 4 to 9 due to the negative squared mass\footnote{The classical 
 solution with $B_{0} = 0$ has no problem, because it has a positive mass 
 unlike the other directions of the field $B$.} of the rank-1 fields 
 $B_{\mu}$. Without restricting the generality, we can focus on the first 
 sphere located in the direction $(x_{1}, x_{2}, x_{3})$, since the three 
 fuzzy spheres all share the same equations of motion.
 
 In the framework of fuzzy 2-spheres, we can solve the equations of motion 
 (\ref{AZeqB+}) and (\ref{AZeqC+}) with the following ansatz  for the rank-2 
 field $C_{ij}$:
  \begin{eqnarray}
   C_{ij} = f(r) \epsilon_{ijk} B_{k}, \label{AZeqC+ansatz}
  \end{eqnarray}
  where $f(r)$ is a function depending on the radius parameter $r$.
  Indeed, the equation of motion (\ref{AZeqC+}) reduces then to:
  \begin{eqnarray}
    \epsilon_{ijk} B_{k} (-f(r) + r + r f^{2}(r) ) =
   0. \label{AZeqC+1} 
  \end{eqnarray}
  (\ref{AZeqC+1}) has two solutions: $f_{\pm} (r) = \frac{1 \pm
  \sqrt{1 - 4r^{2}}}{2r}$. When we plug this result in the equation of motion
  for $B$ (\ref{AZeqB+}), this leads to
  \begin{eqnarray}
   B_{i} ( 1- 2 r f_{\pm}(r) ) = 0.
  \end{eqnarray}
  This gives the same condition on the radius parameter $r$ for 
  both $f_{+}(r)$ and $f_{-}(r)$, namely:
  \begin{eqnarray}
    \sqrt{1 - 4r^{2} } = 0. \label{AZeqBC+so3}
  \end{eqnarray}
  Therefore, when we assume the ansatz (\ref{AZeqC+ansatz}), we obtain 
  the classical solution (\ref{AZso3so3so3}) with the radius parameter 
  set to $r = \frac{1}{2}$, which is fortunately real. Indeed, $r^2 \leq 0$ 
  would indicate that the fuzzy sphere solution is unstable. For example, 
  in the IIB massive matrix model described by (\ref{MIIBMM}), the sign of 
  the squared radius of the fuzzy 2-sphere is linked to the sign of the mass 
  term in the action and it would become negative for a correct-sign mass 
  term, which is to be expected, since in that case, the trivial commutative 
  solution becomes the stable vacuum of the theory.  

  We next want to discuss the stability of the 
  $SO(3) \times SO(3) \times SO(3)$ classical solution in more qualitative
  terms\cite{0101102,0206075}. To this end, we compare the energy of 
  the trivial commutative solution $B_{\mu} = 0$ with that of the
  fuzzy-sphere solution. The classical energy for $B_{\mu} = 0$ is
  obviously $E_{\tiny{ B_{\mu}=0}} = -  S_{\tiny{ B_{\mu}=0}} = 0$.
  \footnote{Recall that in our notation, the action is minus the potential. 
  Since we now consider a classical solution with $B_{0} = 0$ (thus no need 
  of Wick rotation), the energy is simply minus the classical action in which 
  we substitute the solution.} 
  
  In the $SO(3) \times SO(3) \times SO(3)$ fuzzy-sphere background, 
  the 2-form field $C_{ij}$ is
  \begin{eqnarray}
   C_{ij} =  \epsilon_{ijk} B_{k}.  
  \end{eqnarray}
 Therefore, the total energy is
  \begin{eqnarray}
  E_{\tiny SO(3)^3}
  &=& -  S_{\tiny SO(3)^3} = - 64 \mu \sum_{\mu = 1}^{9} Tr(B_{\mu}
  B^{\mu} )
  =  - 3 \times 64 \mu \sum_{i=1}^{3}  Tr(B_{i}
  B_{i}) \nonumber \\ 
  &=& - 12 \mu^{3} N (N-1)(N+1). \label{AZ2so3energy} 
  \end{eqnarray}
 This result shows that the $SO(3) \times SO(3) \times SO(3)$ fuzzy-sphere 
 classical solution has a lower energy compared to the trivial commutative
 solution and hence a higher probability. 

\subsection{Other curved-space solutions and the fuzzy 8-sphere}
 So far, we have analyzed the simplest curved-space solution, the 
 $SO(3) \times SO(3) \times SO(3)$ triple fuzzy spheres. 
 Here, we consider other curved-space classical solutions. 
 The fuzzy $2k$-spheres\cite{9712105,0105006,0111278,0207111},
 which exhibit a $SO(2k+1)$ symmetry, are constructed by the following 
 $n$-fold symmetric tensor product of $(2k+1)$-dimensional gamma matrices:
  \begin{eqnarray}
   B^{SO(2k+1)}_{p} = \frac{\mu r}{2} [ (\Gamma^{(2k)}_{p} \otimes {\bf 1}
   \otimes \cdots \otimes 
   {\bf 1}) + \cdots + ({\bf 1} \otimes \cdots \otimes {\bf 1} \otimes
   \Gamma^{(2k)}_{p} )]_{\textrm{sym}},
  \end{eqnarray}
 where $p$ runs over $1,2, \cdots, 2k+1$. $\Gamma^{(2k)}_{p}$ are $2^{k}
 \times 2^{k}$ gamma matrices, and build a representation of the
 $SO(2k+1)$ Clifford algebra; i.e. $\{ \Gamma^{(2k)}_{p},
 \Gamma^{(2k)}_{q} \} = 2 
 \delta_{pq} {\bf 1}_{2^{k} \times 2^{k}}$. 
 These matrices satisfy the following algebraic relations:
  \begin{eqnarray}
   & &  B^{SO(2k+1)}_{p} B^{SO(2k+1)}_{p} = \frac{\mu^{2} r^{2}}{4} n(n+2k) 
   {\bf 1}_{N_k \times N_k},
   \label{AZ2radius} \\ 
   & & B^{SO(2k+1)}_{pq} B^{SO(2k+1)}_{pq} = 
   - (\frac{\mu r}{2})^{4}  8 k n (n + 2k) {\bf 1}_{N_k \times N_k},
   \label{AZ2radius2} \\      
   & &  [B^{SO(2k+1)}_{pq}, B^{SO(2k+1)}_{s}]  = \mu^{2} r^{2} ( -
     \delta_{ps} B^{SO(2k+1)}_{q} + 
     \delta_{qs} B^{SO(2k+1)}_{p}), \label{AZ2pm} \\
   & &  [B^{SO(2k+1)}_{pq}, B^{SO(2k+1)}_{st}]  =  \mu^{2} r^{2}
     (\delta_{qs} B^{SO(2k+1)}_{pt} + 
      \delta_{pt} B^{SO(2k+1)}_{qs} - \delta_{ps} B^{SO(2k+1)}_{qt} -
     \delta_{qt} B^{SO(2k+1)}_{ps}),  \nonumber \\
     \label{AZ2mm}
  \end{eqnarray}
 where $B^{SO(2k+1)}_{pq} = [B^{SO(2k+1)}_{p},
 B^{SO(2k+1)}_{q}]$ furnishes (up to a normalization factor) a representation
 of the $\mathfrak{so}(2k+1)$ Lie algebra and $N_k$ is the dimension of the
 fully symmetrized irreducible representation for the $SO(2k+1)$ fuzzy sphere.
 The commutation relations (\ref{AZ2pm}) and (\ref{AZ2mm}) are inherited from 
 those of the gamma matrices. Thanks to these relations, we expect that the 
 equations of motion can be solved for all even-dimensional fuzzy spheres 
 in a similar fashion to the fuzzy 2-spheres. In other words, this means that 
 the $SO(2k+1)$ fuzzy spheres will provide us with a whole set of curved
 classical solutions for some precise values of the parameter $r$.
 In addition, the $B^{SO(2k+1)}_{p}$'s satisfy the following self-duality 
 relation:
  \begin{eqnarray}
   \epsilon_{p_{1} \cdots p_{2k+1}} B^{SO(2k+1)}_{p_{1}}
   B^{SO(2k+1)}_{p_{2}} \cdots 
   B^{SO(2k+1)}_{p_{2k}} = \left( \frac{\mu r}{2} \right)^{2k-1} m_{k}
   B^{SO(2k+1)}_{p_{2k+1}}, \label{AZ2duality} 
  \end{eqnarray} 
 where 
  \begin{eqnarray}
   m_{1} = 2i, \hspace{3mm}
   m_{2} = 8(n+2), \hspace{3mm} 
   m_{3} = -48 i (n+2)(n+4), \hspace{3mm}
   m_{4} = - 384 (n+2)(n+4)(n+6), \nonumber \\
   \label{AZ2soksphere} 
  \end{eqnarray}
 which is a generalization of the $\mathfrak{so}(3)$ Lie algebra\footnote{
 It generalizes the fuzzy 2-sphere case where $B_i^{SO(3)}  B_i^{SO(3)}$
is proportional to the identity on the totally symmetric space of dimension
$n+1$. 
 For $N = n+1$, the radius 
 of the $SO(3)$ fuzzy sphere is indeed 
 \begin{eqnarray}
 \frac{\mu^{2} r^{2}}{4} n(n+2) = (\mu r)^{2} \frac{N^{2} -1}{4}. \nonumber
 \end{eqnarray}
 The relation (\ref{AZ2radius}) actually corresponds to the Casimir of the 
 $\mathfrak{so}(3)$ Lie algebra. And (\ref{AZ2duality}) is trivially 
 equivalent to the commutation relation $[B^{SO(3)}_{i}, B^{SO(3)}_{j}] = 
 i \mu r \epsilon_{ijk} B^{SO(3)}_{k}$. \label{ftso3} } and a consequence
 of the duality relation for odd-dimensional Gamma matrices.
 We give the computation of these coefficients in the second appendix.

 Another possible classical solution of our massive
 supermatrix model is the single $SO(9)$ fuzzy sphere. The analysis
 goes in the same way as in the $SO(3) \times SO(3) \times SO(3)$
 fuzzy spheres. Here, the indices $p, q, \cdots$ run over $1, 2,
 \cdots, 9$. For the $SO(9)$ fuzzy-sphere classical solution, we
 likewise assume the following ansatz for the rank-2 fields $C^{SO(9)}_{pq}$: 
  \begin{eqnarray}
   C^{SO(9)}_{pq} = - i \mu^{-1} g(r) B^{SO(9)}_{pq}.
  \end{eqnarray}
 Then, the equation of motion (\ref{AZeqC+}) implies
  \begin{eqnarray}
   \frac{-i}{\mu} B^{SO(9)}_{pq} ( -g(r) + 1 + 7 r^{2} g^{2}(r) ) = 0.
  \end{eqnarray}
 We again have two choices for the function $f(r)$:
  \begin{eqnarray}
   g_{\pm}(r) = \frac{1 \pm \sqrt{1 - 28 r^2}}{14 r^{2}}.
  \end{eqnarray}
 The equation of motion (\ref{AZeqB+}) for the rank-1 field $B^{SO(9)}_{p}$
 gives 
  \begin{eqnarray}
   B^{SO(9)}_{p} ( 1 - 8r^{2} g_{\pm} (r) ) = 0.
  \end{eqnarray}
 Now, unlike the case of the $SO(3) \times SO(3) \times SO(3)$ fuzzy
 spheres, $1 - 8 r^{2} g_{-}(r) = 0$ does not have any real positive
 solution for $r$. However, there is exactly one such solution for $1 - 8r^{2}
 g_{+}(r) = 0$, which is $r = \frac{1}{8}$.

 More generally, for an $SO(2k+1)$ fuzzy sphere, the same ansatz would give
 \begin{eqnarray}
 g_{\pm}(r) &=& \frac{1 \pm \sqrt{1 - 4(2k-1)r^2}}{2(2k-1)r^{2}} \; ,
 \nonumber\\
 1-2kr^{2} g_{\pm} (r)&=& 0 \; , \textrm{ solvable only for } g_{+} (r) 
 \textrm{ at} \\
 r&=&\frac{1}{2k}.\nonumber
  \end{eqnarray}

 We discuss the stability of the $SO(9)$ fuzzy-sphere classical
 solution by computing its classical energy. At the classical level,
 we obtain
  \begin{eqnarray}
   E_{\tiny SO(9)} = - \frac{5}{8} \mu^{3} n(n+8)
   N_4. \label{AZ2so9energy} 
  \end{eqnarray}  
  $N_4$ is given in \cite{0105006}\cite{0111278}
  by\footnote{Generally, $N_k$ is known to be of the order
  ${\cal O}(n^{\frac{k(k+1)}{2}})$, and more explicitly,
   \begin{eqnarray}
    N_1 = (n+1), \hspace{2mm}
    N_2 = \frac{(n+1)(n+2)(n+3)}{6}, \hspace{2mm}
    N_3 = \frac{(n+1)(n+2)(n+3)^{2}(n+4)(n+5)}{360}. \nonumber
   \end{eqnarray}}
   \begin{eqnarray}
   N_4 = \frac{(n+1)(n+2)(n+3)^{2} (n+4)^{2} (n+5)^{2}
   (n+6)(n+7)}{302400}. 
   \end{eqnarray}
  In contrast with the $SO(3) \times SO(3) \times SO(3)$ case, $N$ can
  take here only certain precise 
  values. For example, the smallest non-trivial representation ($n=1$)
  has dimension 16, the following one ($n=2$) 126, then 672, etc...
  The classical energy for the $SO(3) \times SO(3) \times SO(3)$
  fuzzy-sphere solution is of the order ${\cal O}( - \mu^3 n^3 ) = {\cal 
  O}(- \mu^3 N^3)$ while that of the $SO(9)$ fuzzy-sphere solution 
  is of the order ${\cal O}(- \mu^3 n^{12}) = {\cal
  O}( - \mu^3 N^{\frac{6}{5}})$. Therefore, at large $N$, the 
  $SO(3) \times SO(3) \times SO(3)$ triple fuzzy-sphere solution is 
  energetically favored compared to the $SO(9)$ solution at the equal
  size $N$ of 
  the matrices. The presence of a spherical solution for all $N$ in the 
  $SO(3) \times SO(3) \times SO(3)$ case may indeed be a stabilizing
  factor. On the other hand, at equal value of $n$, whose physical meaning 
  is less clear, the fuzzy 8-sphere solution has lower energy.  

  The single $SO(q)$ fuzzy spheres for $q \leq 8$ do not 
  constitute a stable classical solution of our model. When the $SO(q)$ sphere
  occupies the direction $x_{1}, x_{2}, \cdots, x_{q}$, the solution
  $B^{SO(q)}_{q+1} = B^{SO(q)}_{q+2} = \cdots = B^{SO(q)}_{9} =0$ is
  trivially unstable because of the negative mass squared. Whereas, 
  the Cartesian product of several fuzzy spheres, such as $SO(3)
  \times SO(6)$, is a possible candidate for a stable classical solution.  

\subsection{Nucleation process of spherical branes}

Starting from a vacuous spacetime, it is interesting to try to guess how
spherical brane configurations could be successively produced through 
a sequence of decays into energetically more favorable meta-stable brane 
systems. The reader may have noticed that we have so far limited ourselves to 
the study of curved branes building irreducible representations of their
symmetry groups. This could seem at first to be an unjustified prejudice,
but it turns out that such configurations are energetically favored
at equal values of $N$. For example, for $SO(3)$, an irreducible representation
${\cal R}_{N}$ of dimension $N$ contributes as
 \begin{equation}
  E_{{\cal R}_{N}} = - 4 \mu^{3} (N^3-N) \nonumber
 \end{equation}
per fuzzy 2-sphere, while a reducible representation ${\cal R}_{N_1} \oplus 
\ldots \oplus {\cal R}_{N_m}$ of equal dimension $N_1+\ldots+N_m=N$ would
contribute as
 \begin{equation}
  E_{{\cal R}_{N_1} \oplus \ldots \oplus {\cal R}_{N_m}} = - 4 \mu^{3} 
  \sum_{i=1}^{m} (N_i^3-N_i).\nonumber
 \end{equation}
This is obviously a less negative number, especially for big values of $N$.
A similar conclusion was reached in \cite{0101102} for the case of a Euclidean
3-dimensional IIB matrix model with a Chern-Simons term and it seems to be 
a fairly general feature of matrix models admitting non-trivial classical 
solutions. This property is particularly clear for low-dimensional branes, 
since the classical energy is of order ${\cal O}(- \mu^3 N^3)$ for $SO(3)$,
but it remains true for any $SO(2k+1)$ fuzzy-sphere solution, whose energy
is of order ${\cal O}( - \mu^3 N^{1+4/(k(k+1))})$, which also shows
that low-dimensional configurations are favored. As hinted for in the 
preceding subsection, this latter fact can be physically understood by 
remarking that there are more irreps available for low-dimensional fuzzy 
spheres, which makes it easier for them to grow in radius through 
energetically favorable configurations. A third obvious fact is that 
configurations described by representations of high dimensionality are 
preferred.

Put together, these comparisons give us a possible picture for the branes 
nucleation process in this and similar matrix models. As they appear, 
configurations of all spacetime dimensions described by small representations
will be progressively absorbed by bigger representations to form irreducible 
ones, that will slowly grow in this way to bigger values of $N$.
Parallel to that, branes of higher dimensionalities will tend to decay into a 
bunch of branes of smaller dimensionalities, finally leaving only
2-spheres and noncommutative tori of growing radii. If the size of the
Hermitian matrices is left open, as is usually the case in completely
reduced models, where the path integration contains a sum on that
size, no configuration will be truly stable, since the size of the
irreps will grow continuously. 

Of course, this is a relatively qualitative study, which could only be proven
correct by a full quantum statistical study of the model. However, it seems 
to be an interesting proposal for the possible physics of such theories.

\subsection{Supersymmetry}
 We next comment on the structure of the supersymmetry.
 The biggest difference with the purely cubic supermatrix model, due to
 the addition of the mass term, is that this model is not invariant 
 under the inhomogeneous supersymmetry
  \begin{eqnarray}
    \delta_{\tiny \textrm{inhomogeneous}} m=0, \hspace{3mm}
    \delta_{\tiny \textrm{inhomogeneous}} \psi = \xi, \label{AZinhomo}
  \end{eqnarray}
 which is a translation of the fermionic field.
 However, this model has 2 homogeneous supersymmetries in 10 dimensions, 
 which are part of the $\mathfrak{osp}(1|32,\mathbb{R})$ symmetry:
   \begin{eqnarray}
    \delta_{\epsilon} M = \left[ \left( \begin{array}{cc} 0 & \epsilon \\ i 
    {\bar \epsilon} & 0 \end{array} \right), 
     \left( \begin{array}{cc} m & \psi \\ i 
    {\bar \psi} & 0 \end{array} \right) \right]
  = \left( \begin{array}{cc} i (\epsilon {\bar \psi} - \psi {\bar
    \epsilon} )   & - m \epsilon \\ i {\bar \epsilon} m & 0
    \end{array} \right), \label{AZhomo}
   \end{eqnarray}
 which transforms the bosonic and fermionic fields as
  \begin{eqnarray}
   \delta_{\epsilon} m = i (\epsilon {\bar \psi} - \psi {\bar
    \epsilon} ), \hspace{3mm}
   \delta_{\epsilon} \psi = - m \epsilon. \label{AZ3SUSY}
  \end{eqnarray}

  In the IIB matrix model, the supersymmetry has to balance between a quartic 
  term $Tr([A_{\mu},A_{\nu}])^2$ and a trilinear contribution 
  $Tr {\bar \psi} \Gamma^{\mu} [A_{\mu}, \psi]$ in the action 
  (\ref{AZ1IKKT}), which implies that the SUSY transformation of the 
  fermionic field has to be bilinear in the bosonic field. On the other hand, 
  the homogeneous supersymmetries are all linear in the fields in the purely 
  cubic supermatrix model\cite{0102168,0103003}. By incorporating the mass 
  term, we are allowed to integrate out the rank-2 field $C_{\mu_{1} \mu_{2}}$
  by solving the classical equation of motion iteratively as
  in~\cite{0201183}\footnote{The Yang-Mills-like structure of the
  homogeneous SUSY transformation on the fermion comes from $-
  \frac{1}{2} C_{\mu \nu} \Gamma^{\mu \nu} \epsilon$, which is a part
  of $\delta_{\epsilon} \psi = - m \epsilon$. The explicit form of the 
  iterative solution of the equations of motion (\ref{AZeqB+}) and
  (\ref{AZeqC+}) is 
 \begin{eqnarray}
    C_{\mu \nu} &=& - i \mu^{-1} [B_{\mu}, B_{\nu}] 
   + i \mu^{-3} [[B_{\mu}, B_{\rho}], [B_{\nu}, B^{\rho}]] \nonumber \\
   & & - i \mu^{-5} [[B_{\mu}, B_{\rho}], [[B_{\nu}, B_{\chi}],[B^{\rho},
   B^{\chi}]] ]
   + i \mu^{-5} [[B_{\nu}, B_{\rho}], [[B_{\mu}, B_{\chi}],[B^{\rho},
   B^{\chi}]] ] 
   + {\cal O}(\mu^{-7}). \nonumber
 \end{eqnarray}
 }.   

 Thanks to this procedure, the homogeneous SUSY transformation for the 
  fermionic field becomes
  \begin{eqnarray}
  \delta_{\epsilon} \psi = \frac{i}{2} [B_{\mu_{1}}, B_{\mu_{2}}]
  \Gamma^{\mu_{1} \mu_{2}} \epsilon + \cdots, \label{AZ2SUSYferm}
  \end{eqnarray}
  while the transformation of the field $B_{\mu}$ is
  \begin{eqnarray}
   \delta_{\epsilon} B_{\mu} = - \frac{1}{32} tr_{32 \times 32}
      ( i (\epsilon {\bar \psi} - \psi {\bar \epsilon}) \Gamma_{\mu
      \sharp} ) = - \frac{i}{16} {\bar \epsilon} \Gamma_{\mu \sharp}
      \psi. \label{AZ2SUSYvec}
   \end{eqnarray}

  In that sense, the mass term is essential to realize the Yang-Mills-like 
  structure for the homogeneous supersymmetries. On the other hand, if we
  want to preserve the homogeneous supersymmetries, we cannot just put 
  a mass term for $C_{\mu_{1} \mu_{2}}$ by hand, the 
  $\mathfrak{osp}(1|32,\mathbb{R})$
  symmetry forces all fields to share the same mass, since they all
  lie in the same multiplet. In particular, we are forced to introduce a 
  mass term for the fermions as well, which breaks the inhomogeneous
  supersymmetries. In other words, it seems difficult to have Super 
  Yang-Mills-type structure for both homogeneous and inhomogeneous
  supersymmetries in the context of supermatrix models.
  
  Indeed, in contrast with the purely cubic supermatrix 
  model\cite{0102168,0103003}, which has twice as many SUSY parameters,
  the massive supermatrix model has only ${\cal N}=2$ SUSY in 10 dimensions, 
  because it lacks the inhomogeneous supersymmetries. 
  In consequence, we cannot realize the translation of the vector field 
  $A_{\mu}$ as a commutator of two linear combinations of the homogeneous 
  and inhomogeneous supersymmetries (\ref{AZ3SUSY}) as in the IIB matrix model,
  where it leads to the interpretation of the eigenvalues of $A_{\mu}$ as 
  spacetime coordinates. On the contrary,  
   \begin{eqnarray}
   [\delta_{\epsilon}, \delta_{\chi}] m = i [(\epsilon {\bar \chi} -
   \chi {\bar \epsilon}), m], \hspace{3mm}
   [\delta_{\epsilon}, \delta_{\chi}] \psi = i (\epsilon {\bar \chi} - 
   \chi {\bar \epsilon} ) \psi
   \end{eqnarray}
  vanishes up to an $\mathfrak{sp}(32,\mathbb{R})$ rotation. This problem 
  is a serious obstacle for the identification 
  of the supersymmetry of this model with that of the IIB matrix model. 
  More analysis will be reported elsewhere.
 
\section{Conclusion and outlook}
 In this paper, we have investigated a supermatrix model based on
 $\mathfrak{osp}(1|32,\mathbb{R})$ with a mass term and a cubic interaction. 
 To be able to describe the gravitational interaction in terms of large 
 $N$ reduced models, we must understand how the reduced models can describe 
 physics in curved spacetimes. Although the IIB matrix model only possesses 
 flat noncommutative spacetime as a classical solution, by adding a tachyonic 
 mass term as in \cite{0103192}, we can obtain new classical solutions
 building curved space backgrounds. Following this idea, we have expected 
 that massive supermatrix models could also exhibit similar properties 
 leading to non-trivial classical solutions. In particular, 
 we have investigated fuzzy-sphere solutions with symmetries 
 $SO(3) \times SO(3) \times SO(3)$ and $SO(9)$, and calculated the parameter 
 determining the quantization step separating the radii of configurations
 described by different representations of the Lie algebra. We have then 
 discussed their respective likelihood by comparing their energy at the 
 classical level, which gave us a way to understand a possible dynamical
 evolution of the solutions through successively more favorable brane
 configurations.

 It is an intriguing issue to search for other stable curved-space
 classical solutions. For example, an $SO(3) \times SO(6)$ fuzzy sphere
 could be a promising candidate. Indeed, the expansion around this 
 classical solution may be related in some way to the BMN matrix
 model\cite{0202021,0205185,0206239,0207190}, which appears as the discrete
 light-cone quantization of D0-brane in the M-theory pp-wave
 background\cite{0202021}. However, to study this case explicitly, 
 we first have to analyze how the equations of motion
 (\ref{AZeqB+}) and (\ref{AZeqC+}) can be treated in the case of odd
 fuzzy spheres. 
 Indeed, as is outlined in \cite{0207111}, the commutator $[B_i,B_j]$ does not 
 correspond to a representation of $SO(2k)$ for odd fuzzy spheres, which 
 makes the analysis much more involved. However, the construction of an
 $SO(3) \times SO(6)$ classical solution would show how transverse 5-branes 
 can appear in this model. The $SO(4) \times SO(5)$ case should proceed along
 similar lines. Another case that can be investigated is a solution of the
 type $SO(2) \times SO(2) \times SO(5)$, in which the two first circles build
 a noncommutative torus as in \cite{0103192}.

 Another difficult problem that could be tackled in the future is a detailed
 analysis of the stability under perturbations of such noncommutative spaces 
 in this and other similar models, to see if these solutions are really local 
 minima of the potential energy.
  
 We next mention the relation between our massive 
 $\mathfrak{osp}(1|32,\mathbb{R})$ supermatrix model and the IIB matrix 
 model. In the quantum field theory, some different models which possess 
 the same symmetry are equivalent in the continuum limit. This property is 
 known as universality. We expect that some similar mechanism may hold true 
 of the large $N$ reduced models, and hence that various matrix models may 
 have the same large $N$ limit. If we believe in the matrix-model version of 
 the universality conjecture, it is possible that our massive supermatrix 
 model could be equivalent to the IIB matrix model.
 In this sense, it is interesting to investigate further whether our
 model shares the maximal 10-dimensional ${\cal N}=2$ SUSY with the IIB
 matrix model and how.   

 \paragraph{Acknowledgment} \hspace{0mm} \\
 The authors would like to express their gratitude to  Luca Carlevaro, 
 Jean-Pierre Derendinger, Frank Ferrari, Hikaru Kawai, Yusuke Kimura and 
 Christian R\"omelsberger for valuable discussion.
 The works of T.A. were supported in part by Grant-in-Aid for
 Scientific Research from Ministry of Education, Culture, Sports,
 Science and Technology of Japan ($\#$01282). M.B. acknowledges the financial 
 support provided through the European Community's Human Potential Programme 
 under contract HPRN-CT-2000-00131 Quantum Spacetime and the Swiss Office for 
 Education and Science as well as the Swiss National Science Foundation.

\section{Massive supermatrix model action}
\begin{eqnarray}
   S &=& 96 \mu Tr \left( - W^{2} - A_{\mu} A^{\mu} + B_{\mu} B^{\mu}
   + \frac{1}{2} C_{\mu_{1} \mu_{2}} C^{\mu_{1} \mu_{2}}
     - \frac{1}{4!} H_{\mu_{1} \cdots \mu_{4}} H^{\mu_{1} \cdots
     \mu_{4}} \right. \nonumber \\
    & & \hspace{15mm} \left. 
     - \frac{1}{5!} Z_{\mu_{1} \cdots \mu_{5}} Z^{\mu_{1} \cdots
     \mu_{5}} 
     +  \frac{i}{16} {\bar \psi} \psi \right) \nonumber \\
    &+& 32 i Tr \Bigg( - 3 C_{\mu_{1} \mu_{2}} [A^{\mu_{1}},
     A^{\mu_{2}}] + 3 C_{\mu_{1} \mu_{2}}  [B^{\mu_{1}}, B^{\mu_{2}}] 
     + 6 W [A_{\mu}, B^{\mu}]
     + C_{\mu_{1} \mu_{2}} [ {C^{\mu_{2}}}_{\mu_{3}}, C^{\mu_{3}
     \mu_{1}}] \nonumber \\
   & & \hspace{12mm} + \frac{1}{4} B_{\mu_{1}} [H_{\mu_{2} \cdots
     \mu_{5}}, Z^{\mu_{1} \cdots \mu_{5}}]
     - \frac{1}{8} C_{\mu_{1} \mu_{2}}
     ( 4 [ {H^{\mu_{1}}}_{\rho_{1} \rho_{2} \rho_{3}}, H^{\mu_{2}
     \rho_{1} \rho_{2} \rho_{3}}]
      + [ {Z^{\mu_{1}}}_{\rho_{1} \cdots \rho_{4}}, Z^{\mu_{2}
     \rho_{1} \cdots \rho_{4}}] ) \nonumber \\
   & & \hspace{12mm} + \frac{3}{(5!)^{2}} \epsilon^{\mu_{1} \cdots
     \mu_{10} \sharp} \left( - W [Z_{\mu_{1} \cdots \mu_{5}},
     Z_{\mu_{6} \cdots \mu_{10}}] + 10 A_{\mu_{1}} [H_{\mu_{2} \cdots
     \mu_{5}}, Z_{\mu_{6} \cdots \mu_{10}}] \right) \nonumber \\
   & & \hspace{12mm} + \frac{200}{(5!)^{3}}
     \epsilon^{\mu_{1} \cdots  \mu_{10} \sharp}
     \left( 5 H_{\mu_{1} \cdots \mu_{4}}  [{Z_{\mu_{5} \mu_{6} \mu_{7}}}^{\rho
     \chi}, Z_{\mu_{8} \mu_{9} \mu_{10} \rho \chi}]
      +10 H_{\mu_{1} \cdots \mu_{4}} [ {H_{\mu_{5} \mu_{6}
     \mu_{7}}}^{\rho}, H_{\mu_{8} \mu_{9} 
     \mu_{10} \rho}]  \right. \nonumber \\
   & & \hspace{35mm} \left.  + 6 {H^{\rho \chi}}_{\mu_{1} \mu_{2}} 
     [ Z_{\mu_{3} \mu_{4} \mu_{5} \rho \chi}, Z_{\mu_{6} \cdots
     \mu_{10}} ] \right) \Bigg)\nonumber \\
   &+& 3 Tr \left( {\bar \psi} \Gamma^{\sharp} [W, \psi]
         + {\bar \psi} \Gamma^{\mu} [A_{\mu}, \psi]
         + {\bar \psi} \Gamma^{\mu \sharp} [B_{\mu}, \psi]
         + \frac{1}{2!} {\bar \psi} \Gamma^{\mu_{1} \mu_{2}}
           [C_{\mu_{1} \mu_{2}}, \psi] \right. \nonumber \\
   & & \hspace{10mm} + \left. \frac{1}{4!} {\bar \psi} \Gamma^{\mu_{1} \cdots 
     \mu_{4} \sharp} [H_{\mu_{1} \cdots \mu_{4}}, \psi] 
     + \frac{1}{5!} {\bar \psi} \Gamma^{\mu_{1} \cdots \mu_{5}}
      [Z_{\mu_{1} \cdots \mu_{5}}, \psi] \right)
  \end{eqnarray} 
 \section{Notations and useful formulae} 
 In this appendix, we give hints for the derivations of the coefficients 
 $m_{k}$ in the self-duality relation (\ref{AZ2duality}) for the $SO(2k+1)$
 fuzzy sphere. We first define the $2^{k} \times 2^{k}$
 gamma matrices in the $2k$-dimensional Euclidean space
 $\Gamma^{(2k)}_{p}$ by the following recursive relation:
  \begin{eqnarray}
 & &  \Gamma^{(2k+2)}_{p} = \Gamma^{(2k)}_{p} \otimes \sigma_{2} = 
   \left( \begin{array}{cc} 0 & -i \Gamma^{(2k)}_{p} \\ i
   \Gamma^{(2k)}_{p} & 0 \end{array} \right),
   \hspace{3mm} 
   \Gamma^{(2k+2)}_{2k+2} = {\bf 1}_{2^{k} \times 2^{k}} \otimes
   \sigma_{1} = \left( \begin{array}{cc} 0 & {\bf 1}_{2^{k} \times
   2^{k}} \\ {\bf 1}_{2^{k} \times 2^{k}} & 0 \end{array} \right), 
   \nonumber \\
 & & \Gamma^{(2k+2)}_{2k+3} = {\bf 1}_{2^{k} \times 2^{k}} \otimes
   \sigma_{3} = \left( \begin{array}{cc} {\bf 1}_{2^{k}
   \times 2^{k}} & 0 \\ 0 & - {\bf 1}_{2^{k} \times 2^{k}}
   \end{array} \right), \label{AZCrecur}
  \end{eqnarray}
 where the index $p$ runs over $p=1,2, \cdots, 2k+1$. The
 2-dimensional gamma matrices are identical to the Pauli matrices:
 $\Gamma^{(2)}_{i} = \sigma_{i}$. Under this notation, we obtain 
  \begin{eqnarray}
   \sigma_{1} \sigma_{2} = i \sigma_{3}, \hspace{3mm} 
   \Gamma^{(4)}_{1} \Gamma^{(4)}_{2} \Gamma^{(4)}_{3} \Gamma^{(4)}_{4} 
   = \Gamma^{(4)}_{5}, \hspace{3mm}
   \Gamma^{(6)}_{1} \Gamma^{(6)}_{2} \cdots \Gamma^{(6)}_{6} = - i
   \Gamma^{(6)}_{7}, \hspace{3mm}  
   \Gamma^{(8)}_{1} \Gamma^{(8)}_{2} \cdots \Gamma^{(8)}_{8} = -
   \Gamma^{(8)}_{9}. \nonumber \\ \label{AZCchirality}
  \end{eqnarray} 
 It is trivial that $m_{1} = 2i$ as explained in the footnote \ref{ftso3},
 while the computation of the coefficient $m_{2}$ can be found in
 \cite{9712105}. In this appendix, we give formulae that are useful to derive
 $m_{3}$ and $m_{4}$. We set $\frac{\mu r}{2} = 1$ and omit 
 "$_{\textrm{sym}}$", with the understanding that these formulae are only 
 valid in the fully symmetrized representations. 
 
 In general, we have
  \begin{eqnarray}
 & &\sum_{l=1}^{2k+1}  (\Gamma^{(2k)}_{l} \otimes \Gamma^{(2k)}_{l}) = 
  ({\bf 1}_{2^k \times 2^k} \otimes {\bf 1}_{2^k \times 2^k}), \nonumber \\
 & &\sum_{l=1}^{2k+1}  (\Gamma^{(2k)}_{l_{1} l_{2}} \otimes
   \Gamma^{(2k)}_{l_{1} l_{2}}) = -2k ({\bf 1}_{2^k \times 2^k} \otimes 
   {\bf 1}_{2^k \times 2^k})
  \end{eqnarray}
 More specifically, to compute $m_4$, we also need
\begin{eqnarray}
 & & (\Gamma^{(6)}_{l_{1} l_{2} l_{3}} \otimes 
   \Gamma^{(6)}_{l_{1} l_{2} l_{3}} ) = 
   - 18 ({\bf 1}_{8 \times 8} \otimes {\bf 1}_{8 \times
   8}), \label{AZCfreq43} \\
 & & (\Gamma^{(6)}_{l_{1} l_{2}}
   \otimes \Gamma^{(6)}_{l_{3}} \otimes \Gamma^{(6)}_{l_{1}
   l_{2} l_{3}}) = -6  ({\bf 1}_{8 \times 8} \otimes {\bf 1}_{8 \times 
   8} \otimes {\bf 1}_{8 \times 8}), \label{AZCfreq44} \\
 & &  (\Gamma^{(6)}_{l_{1} l_{2}}
   \otimes \Gamma^{(6)}_{l_{3} l_{4}} \otimes \Gamma^{(6)}_{l_{1}
   \cdots l_{4}}) = 24 ({\bf 1}_{8 \times 8} \otimes {\bf 1}_{8 \times 
   8} \otimes {\bf 1}_{8 \times 8}), \label{AZCfreq45} \\
 & & (\Gamma^{(6)}_{l_{1} l_{2}} \otimes \Gamma^{(6)}_{l_{3} l_{4}}
   \otimes \Gamma^{(6)}_{l_{5} l_{6}} \otimes \Gamma^{(6)}_{l_{1}
   \cdots l_{6}} )
 = - 48 ({\bf 1}_{8 \times 8} \otimes {\bf 1}_{8 \times 8} \otimes {\bf
   1}_{8 \times 8} \otimes {\bf 1}_{8 \times 8}). \label{AZCfreq46}
 \end{eqnarray}

\end{document}